\documentclass[aps,prapplied,reprint,superscriptaddress]{revtex4-2}

\usepackage[colorlinks=true,citecolor=blue,urlcolor=black]{hyperref}
\usepackage{qip}
\usepackage{amsmath,amssymb}
\usepackage{amsfonts}
\usepackage{braket}
\usepackage{mathtools}
\usepackage{nccmath}
\usepackage{siunitx}
\usepackage{xcolor}
\usepackage[english]{babel}
\usepackage[utf8]{inputenc}
\inputencoding{utf8}

\DeclarePairedDelimiter\abs{\lvert}{\rvert}%

\DeclareMathOperator{\etadet}{\eta_{\text{det}}}
\newcommand{\norm}[1]{\left\lVert#1\right\rVert}

\newcommand{\Z}{\mathsf{Z}}
\newcommand{\X}{\mathsf{X}}
\newcommand{\Y}{\mathsf{Y}}
\newcommand{\Sset}{\mathcal{S}}

\newcommand{\be}{\begin{equation}}
\newcommand{\ee}{\end{equation}}
\usepackage{scalerel,stackengine}
\stackMath
\newcommand\widecheck[1]{
\savestack{\tmpbox}{\stretchto{
  \scaleto{
    \scalerel*[\widthof{\ensuremath{#1}}]{\kern-.6pt\bigwedge\kern-.6pt}
    {\rule[-\textheight/2]{1ex}{\textheight}}
  }{\textheight}
}{0.5ex}}
\stackon[1pt]{#1}{\scalebox{-1}{\tmpbox}}
}

\begin{document}

\title{Estimating the Photon-Number Distribution of Photonic Channels for Realistic Devices and Applications in Photonic Quantum Information Processing}

\author{Emilien \surname{Lavie}}
\email{emilien.lavie@u.nus.edu}
\affiliation{Department of Electrical \& Computer Engineering, National University of Singapore, Singapore }
\author{Ignatius William Primaatmaja}
\affiliation{Centre for Quantum Technologies, National University of Singapore, Singapore}
\author{Wen Yu \surname{Kon}}
\affiliation{Department of Electrical \& Computer Engineering, National University of Singapore, Singapore }
\author{Chao \surname{Wang}}
\affiliation{Department of Electrical \& Computer Engineering, National University of Singapore, Singapore }
\author{Charles Ci Wen \surname{Lim}}
\affiliation{Department of Electrical \& Computer Engineering, National University of Singapore, Singapore }
\affiliation{Centre for Quantum Technologies, National University of Singapore, Singapore}

\begin{abstract}

Characterising the input-output photon-number distribution of an unknown optical quantum channel is an important task for many applications in quantum information processing.
Ideally, this would require deterministic photon-number sources and photon-number-resolving detectors, but these technologies are still work-in-progress. In this work, we propose a general method to rigorously bound the input-output photon number distribution of an unknown optical channel using standard optical devices such as coherent light sources and non-photon-number-resolving detectors/homodyne detectors. To demonstrate the broad utility of our method, we consider the security analysis of practical quantum key distribution systems based on \emph{calibrated} single-photon detectors and an experimental proposal to implement time-correlated single photon counting technology using homodyne detectors instead of single-photon detectors.

\end{abstract}

\maketitle

\section{Introduction}
Quantum photonics is the art of using low-light optical signals to exchange and process information in the quantum regime~\cite{flamini_photonic_2018, slussarenko_photonic_2019}.
Today, photonic systems represent one of the most promising platforms to implement quantum technology, with already several well-established applications ranging from quantum cryptography~\cite{pirandola_advances_2020, xu_secure_2020} and communications~\cite{gisin_quantum_2007}, to sensing and metrology~\cite{giovannetti_quantum-enhanced_2004,giovannetti_quantum_2006,giovannetti_advances_2011}, lithography \cite{boto_quantum_2000} and imaging~\cite{shih_quantum_2007}.

\begin{figure}[t!]
\includegraphics[width=\columnwidth]{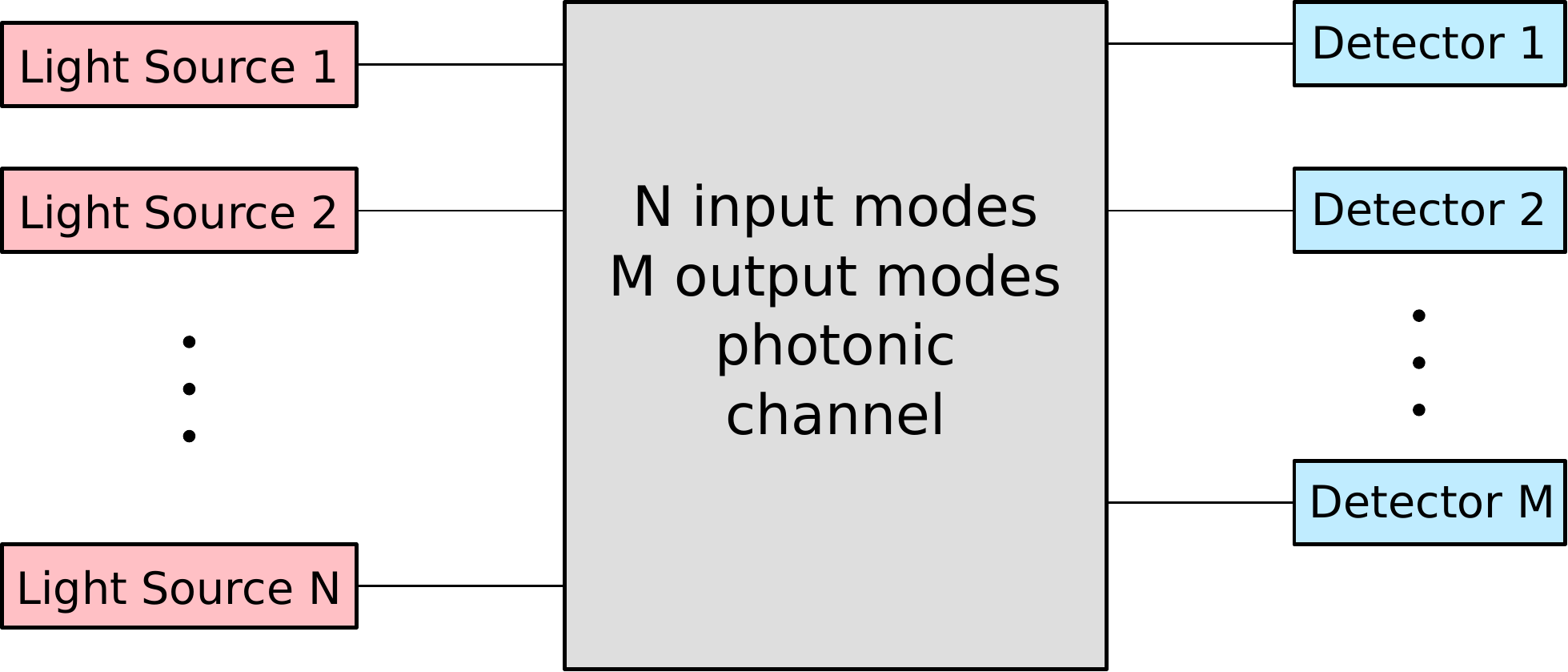}
\caption{Typical scheme for estimating the input-output photon-number distribution $q(\vec{m}|\vec{n})$ of a photonic channel. Each of the $N$ input modes of the photonic channel is connected to a light source that supplies the input photons and each of the $M$ output modes is connected to a detector that gives an outcome related to the number of photons leaving the photonic channel. Note that the sources and detectors considered here may include some form of active modulation devices, such as intensity modulators.
}
\label{fig:mimo}
\end{figure}

In the most general setting, one considers the preparation, transmission, and detection of optical signals.
Here, the photonic channel of interest accepts an $N$-mode input state and returns an $M$-mode output state, which is then measured by a series of photon-counting devices.
More formally, let $q(\vec{m}|\vec{n})$ be the probability of obtaining $\vec{m}=(m_1,m_2,\ldots,m_M)$ photons across the $M$ output modes given $\vec{n}=(n_1,n_2,\ldots,n_N)$ photons are injected into the channel (see Figure~\ref{fig:mimo}).
Here, we note $q(\vec{m}|\vec{n})$ can be characterised independently of the channel's dynamics.
That is, the knowledge of the channel is not needed to estimate $q({\vec{m}|\vec{n}})$, i.e., we can treat the channel as a black box with $N$ inputs and $M$ outputs and sample accordingly.
To keep the characterisation as general as possible, we also allow cases in which the channel is not photon-number preserving.
This can happen when the channel experiences loss and/or suffers from background noise.
Correspondingly, if the channel is known to be photon-number preserving, then we have $\sum_{i=1}^Nn_i = \sum_{j=1}^Mm_j$.

In practice, the input-output photon-number distribution is central to a multitude of information processing tasks.
A good example is boson-sampling---a type of non-universal quantum computation~\cite{aaronson_computational_2011,brod_photonic_2019}.
Here, a linear optical interferometer with $N$ inputs and $N$ outputs is considered, where the input is injected with a fixed number of single photons and the output measured with photon-counting devices.
In Ref.~\cite{aaronson_computational_2011}, it was shown that $q(\vec{m}|\vec{n})$ evaluates directly the permanents of sub-matrices of the interferometer's matrix.
On the other hand, solving such matrix permanents with a classical computer is known to be computationally hard.

Another example is quantum key distribution (QKD), particularly those using discrete variable encoding~\cite{bennett_quantum_2014,scarani_security_2009}.
For such a communication system (connected by a quantum channel with one input and one output), the estimation of single-photon statistics is essential for protocol security.
Take for instance $q(1|1)$, which quantifies how often the untrusted channel behaves as a true single-photon channel.
Having this information strengthens the QKD security analysis by allowing one to assume that (1) the adversary forwards exactly one photon to the receiver and (2) the trusted detector noise and untrusted channel noise are separated.
The former is especially powerful as it enables the security analysis of practical QKD under the assumption of a qubit channel (it also applies generally to any qudit channel of interest and hence to high-dimensional QKD as well~\cite{cerf_security_2002,islam_securing_2018,islam_scalable_2019}).

As a final example, we consider time-correlated single-photon counting (TCSPC) \cite{becker_overview_2005, oconnor_basic_1984}, an optical waveform measurement technique that is widely used in fundamental physics research (e.g., ranging \cite{ren_laser_2011, mccarthy_kilometer-range_2013}, imaging \cite{pawlikowska_single-photon_2017, tobin_three-dimensional_2019}, light source characterisation, and life sciences experiments (e.g., fluorescence-lifetime imaging microscopy \cite{yguerabide_nanosecond_1972}
).
In this setting, the input and output optical modes of the channel are temporal modes corresponding to different time intervals.
Here, we have $\sum_{i=1}^Nn_i =1 $ and $\sum_{j=1}^Mm_j \leq 1$ since only one photon is deployed in a single trial.
In recent years, to improve the efficiency and speed of optical waveform measurement, the idea of using photon-number resolving measurements has been proposed as well \cite{li_time-correlated_2016, dai_realization_2020}.

The input-output photon-number distribution can be easily estimated if one injects fixed photon-number states $\ket{n_1}\ket{n_2}\ldots\ket{n_N}$ into the channel and measure the output using photon-counting devices (which counts the number of photons in each output mode).
However, this would require deterministic photon-number sources and photon-number resolving detectors (PNRDs), which at the moment are still in development~\cite{flamini_photonic_2018,slussarenko_photonic_2019}.
The more realistic options are probabilistic photon-number sources (e.g., coherent lasers) and non-photon-number resolving detectors (e.g., threshold detectors and homodyne detectors); these optical devices are not only highly reliable and cost-effective but also widely available.

To this end, it is natural to ask if one can use these standard optical devices to estimate the input-output photon-number distribution of an unknown photonic channel.

It is worthwhile to mention that the problem we are interested in can be seen as a special case of \emph{coherent-state quantum process tomography} (csQPT), which uses coherent states (probe states) and homodyne measurement to reconstruct the process matrix of an unknown optical channel~\cite{lobino_complete_2008, rahimi-keshari_quantum_2011, anis_maximum-likelihood_2012, fedorov_tomography_2015}.
Indeed, by looking at only the diagonal components of the process matrix, one can recover the photon-number distribution of the unknown channel.
Importantly, since we are only interested in the photon-number distribution, the implementation can be significantly simplified, i.e., the probe states and local oscillators can come from independent laser sources, as we will show later.
Note that in the case of csQPT, the relative phase between the probe states and the local oscillator has to be calibrated (needed to fully recover the underlying process matrix), which may be an issue in long-distance quantum communication protocols such as QKD~\cite{qi_generating_2015,soh_self-referenced_2015}.

The problem of estimating the photon-number distribution of an unknown optical channel is not new and has been studied before in the field of QKD using threshold detectors. On the input side, \emph{decoy-state method} has been proposed, which uses phase-randomised light pulses with different intensities to estimate single-photon statistics~\cite{hwang_quantum_2003,lo_decoy_2005,wang_beating_2005}.
The method essentially entails solving a system of linear equations constrained by the different expected detection rates of the protocol.
As such, there are two approaches towards solving the problem, namely one can do it analytically via Gaussian elimination~\cite{ma_practical_2005,tsurumaru_exact_2008} or numerically with linear programming~\cite{ma_statistical_2012,curty_finite-key_2014}.
The same principle can also be applied to threshold detectors to estimate the output photon-number distribution of the channel~\cite{moroder_detector_2009}.
In this approach, called \emph{detector-decoy method}, one randomly varies the detection efficiency with a variable optical attenuator or intensity modulator to generate a system of linear equations; likewise, these are constrained by the different detection rates effected by the  variation of detection efficiency. Given that both decoy-state and detector-decoy methods are based on the same concept, it is thus natural to consider the combination of these two approaches.
This direction was recently pursued by the authors of Ref.~\cite{navarrete_characterizing_2018}, who used the direct combination of decoy-state and detector-decoy methods to characterise multi-photon quantum interference patterns. Alternatively, the authors of Ref.~\cite{zhang_generalized_2020} used a source modulation along with PNRDs to characterise multi-photon quantum interference.

Here, based on the above ideas, we provide a systematic approach to analyse optical communication systems using a linear estimation of the photon-number statistics, extending the decoy-state, detector-decoy and homodyne based linear estimation methods. Our theoretical contributions are three-fold: (1) the extension of Ref.~\cite{navarrete_characterizing_2018} to homodyne detectors, (2) the security analysis of practical QKD systems based on \emph{calibrated} single-photon detectors, and (3) an experimental proposal to implement TCSPC technology using homodyne detectors instead of single-photon detectors.
Concerning the latter, there are two practical advantages in using homodyne detectors: (1) these detectors are typically much more cost-effective than single-photon detectors and (2) no active intensity modulation is required to achieve the same effect as detector-decoy.
More generally, our extended approach with homodyne detectors provides a simpler and more cost-effective implementation path for applications that requires only the knowledge of $q(\vec{m}|\vec{n})$ instead of single-shot information (see the examples above).
We also present two different methods to estimate the desired input-output photon-number probabilities: one based on Gaussian elimination and the other based on linear programming.

Additionally, we highlight that our work is focused on providing interval estimates on the photon-number statistics instead of point estimation.
Indeed, our approach is essentially motivated by how parameters are estimated in QKD: there, it is imperative to provide reliable upper and lower bounds on parameters such as bit error rates and detection rates, which characterise the amount of key information leaked to the unknown channel.
Therefore, the methods that we describe later include a guarantee on the statistical distance to the true value, unlike other estimation techniques such as maximum-likelihood~\cite{banaszek_maximum-likelihood_1998} and least square estimation~\cite{tan_inverse_1997}.

The paper is organised as follows.
In Section~\ref{sec:models}, we introduce a general channel model and the optical device models used in the estimation.
Then in Section~\ref{sec:methods}, we present two methods for bounding the desired photon-number probabilities. Finally, in Section~\ref{sec:applications} we show how our method can be used to analyse the security of practical QKD with calibrated detectors and TCSPC using homodyne detectors.

\section{Channel Modelling}
\label{sec:models}
In the following, we keep our analysis to a single-mode photonic channel (i.e. a channel with one input mode and one output mode); the generalisation to multi-mode channels is straightforward.
The starting point of our approach is the measurement function, $\{f(x,y)\}_{x,y}$, which characterises the observed statistics depending on two designated control parameters $x$ and $y$ owned respectively by Alice (transmitter side) and Bob (receiver side).
Here, the parameters are quantities used to test the unknown photonic channel.
In the case of active schemes, they are random optical modulations operated by the users.
In the case of passive schemes (e.g. passive decoy states \cite{mauerer_quantum_2007, curty_non-poissonian_2009, xu_improvement_2009, curty_passive_2010, zhang_simple_2018} or homodyne detection as presented later in this paper), they are random variables whose outcomes are correlated to the behaviour of the unknown channel.

In the most general setting, the measurement function is modelled by
\be\label{eq:meas_dist}
f(x,y)=\sum_{n,m= 0}^{\infty}\underbrace{p_n(x)}_{\rm{transmitter}}\underbrace{q(m|n)}_{\rm{channel}} \underbrace{r_m(y)}_{\rm{receiver}},
\ee
where $p_n(x)$ is the input photon-number distribution (representing correlations between a $n$-photon state transmission event and Alice's parameter $x$), $q(m|n)$ is the probability of the channel emitting $m$ photons given it has received $n$ photons, and $r_m(y)$ is the measurement response representing the correlation between a $m$-photon reception event and Bob's parameter $y$.

As mentioned above, our goal is to estimate certain elements of the unknown channel's input-output photon-number distribution, $q(m|n)$.
To that end, we suppose the input photon-number distribution $p_n(x)$, the measurement response $r_m(y)$, and the measurement function $f(x,y)$ are fully characterised for any $x$ and $y$.
That is, we assume the user has complete knowledge of the underlying optical devices and has made enough measurements to accurately infer $f(x,y)$.

Similar to standard decoy-state method implementations, we use a phase-randomised coherent-wave laser to generate photon-number states at the channel's input.
In this case, the light field entering the channel is described by a Poisson distribution of photon-number states
\be
\rho_{\mu}=\sum_{n\geq 0} \frac{\mu^n e^{-\mu}}{n!}\proj{n},
\ee
where $\mu$ is the mean photon number of the field.
The random input $x$ is achieved by modulating the mean photon number with an intensity modulator.
As such, the probability model of the source is fully characterised by $x$ and given by
\be
\label{eq:laser}
p_n(x)= \frac{x^n}{n!}e^{-x}.
\ee

For the measurement model, we can use either a threshold detector or a homodyne detector.
In the former case, the detector only fires if some photons are detected. As such, there are only two possible outcomes, detection and no detection.
Here, we consider only the no detection outcome since the detection outcome is simply the complement event.
Following Ref.~\cite{moroder_detector_2009}, the response of a practical threshold detector can be modelled using
\be
\label{eq:threshold}
r_m(y=\nu)=(1-p_{\rm{dc}})(1-\nu\etadet)^m,
\ee
where $p_{\rm{dc}}$ is the probability of dark count, $\etadet$ is the single-photon efficiency of the detector, and $\nu$ is the transmission efficiency of the intensity modulator (placed in front of the detector) controlled by input $y$.
We note that this model is general and applies to most of today's standard single-photon detection techniques, e.g., single-photon avalanche diodes (SPADs) and superconducting nano-wire single-photon detectors (SNSPDs); see Ref.~\cite{eisaman_invited_2011} for an overview of single-photon technology.
This model is simple and might be inaccurate under specific operating conditions like fast repetition rate under which other effects like after-pulses might appear.
However it is easy to replace the simple model used in Eq.~\eqref{eq:threshold} by a more refined model like the one suggested in \cite{fan-yuan_afterpulse_2018} to account for such effects.
For simplicity in this paper, we stick to the simple model to avoid unnecessary complication in the understanding of the underlying method.

In the case of homodyne detection, it does not count the number of photons in the incoming light field but rather gives an outcome whose probability density function is correlated to the number of photons~\cite{shapiro_quantum_1985}.
This relation becomes more apparent when the local oscillator is phase-randomised and the response of the detector when given  $m$ photons is given by~\cite{tan_inverse_1997, banaszek_maximum-likelihood_1998}:
\be
\label{eq:homodyne}
r_m(y) =\sum^m_{k=0} {m \choose k} \frac{\etadet^k \big(1-\etadet\big)^{m-k}}{\sqrt{\pi}2^k k!}H^2_k(y)e^{-y^2},
\ee
where $y$ is a real number and $\{H_k(y)\}_k$ are Hermite polynomials \cite{szego_laguerre_1939, thangavelu_hermite_1993}.

To estimate the desired photon-number distribution, several statistical methods can be employed, e.g., those based on linear estimation \cite{munroe_photon-number_1995, leonhardt_sampling_1996}, least square estimation \cite{tan_inverse_1997}, and maximum-likelihood \cite{banaszek_maximum-likelihood_1998}.

Here, two observations are in order.
Firstly, unlike threshold detectors, one can obtain any number of discrete outcomes by binning  $y$ (in practice, an Analog-to-Digital Converter (ADC) is used).
Secondly, notice that no additional intensity modulation is required here.
This is because the response density function of a homodyne detector is sensitive to the range of $y$ and hence one can optimise the binning function (i.e., the ADC) to assign different weights to different input photon-number states.
Essentially, this is the same as the detector-decoy method, which assigns different detection probabilities to different photon-number states via the variation of the detection efficiency.
Again here, the model we use in Eq.~\eqref{eq:homodyne} is relatively simple and could be refined to include additional imperfection of realistic detectors like electronic noise \cite{appel_electronic_2007}.

\section{Methods}
\label{sec:methods}
In most quantum information processing tasks, one is only interested in elements of $q(\vec{m}|\vec{n})$ that are small in the input photon number and output photon number. Additionally, the possible values for the controlled parameters $x$ and $y$ are limited to fixed sets $x \in \mathcal{X}:= \{x_0,x_1,\ldots,x_{n_0}\}$ and $y \in \mathcal{Y}:= \{y_0,y_1,\ldots,y_{m_0}\}$. We consider a practically-relevant finite subset of $q(m|n)$ by focusing on $n \in \mathcal{N}_0:= \set{0,1, \dots ,n_0}$ and $m \in \mathcal{M}_0:= \set{0,1, \dots ,m_0}$. Our objective is to derive upper and lower bounds on a specific element or a linear combination of different elements from the set $\{q(m|n)\}_{n\in\mathcal{N}_0,m\in\mathcal{M}_0}$. As mentioned, this problem is essentially a linear optimisation problem with constraints given by positivity, normalisation, and the measurement distribution. More specifically, for positivity one has $q(m|n)\geq 0$ for any $n$ and $m$, for sub-normalisation $\sum_{m=0}^{m_0}q(m|n)\leq 1$ for any $n$, and for measurement distribution
\be
f(x,y)\geq \sum_{n=0}^{n_0}\sum_{m=0}^{m_0}p_n(x)q(m|n) r_m(y),
\ee for any $x$ and $y$. In addition, one could also exploit the knowledge of characterised functions $p_n(x)$ and $r_m(y)$ to construct linear constraints like
\be 0\leq f(x,y)-\sum_{n=0}^{n_0}\sum_{m=0}^{m_0}p_n(x)q(m|n) r_m(y) \leq h(x,y),\ee where $h(x,y)$ is some positive function depending on the optical devices used in the application. We will provide some examples later in Section \ref{sec:applications} and more technical details in Appendix \ref{app:bound_derivation}. In the following, we present two methods to estimate the desired input-output photon-number statistics. \newline

\noindent\emph{Linear programming method}: Let $q(m^*|n^*)$ be the quantity of interest to which an upper bound is desired, then the linear programming (LP) problem is
\begin{equation}
\begin{split}
    \rm{Maximise:}\quad &q(m^*|n^*)\\
    \text{subject to:}\quad & 0\leq q(m|n)\leq 1,\, \forall~ n\leq n_{c},\,m\leq m_{c}\\
    &\sum_{m=0}^{m_{c}}q(m|n)\leq 1,\,\forall~n\leq n_{c}\\
    &\sum_{n=0}^{n_c}\sum_{m=0}^{m_c}p_n(x)q(m|n)r_m(y) \\ & \leq f(x,y),\,\forall~x,y\\
    & \sum_{n=0}^{n_c}\sum_{m=0}^{m_c}p_n(x)q(m|n)r_m(y)\\&
    \geq f(x,y)-h(x,y),\,\forall~x,y.
\end{split}
\end{equation}
Evidently, the idea behind LP is to use the various constraints on $q(m|n)$ or some linear combination of them to provide bounds for the possible values of $q(m|n)$.
Also, since the optimisation is numerical, it is useful to first narrow down to a set of $m$ and $n$ of interest, which can be done via the truncation of both $m$ and $n$ up till some suitable choice of $m_{c}\geq m_0$ and $n_{c} \geq n_0$. Notably, LP is performed by first defining a feasible region where the set of input-output distribution $\{q(m|n)\}_{m\leq m_c,n\leq n_c}$ satisfies the constraints.
One can then maximise (resp. minimise) the desired probability $q(m^*|n^*)$ over the feasible region to obtain the upper (resp. lower) bound on $q(m^*|n^*)$.
\newline

\noindent\emph{Analytical method}:
The basic idea of the second method is to leverage the complete knowledge of the characterised devices to estimate $q(m^*|n^*)$ without using any cutoff condition $n\leq n_c$ or $m\leq m_c$.
To that end, we consider a linear combination of the measurement function (see Eq.~\eqref{eq:meas_dist}) over a finite set of evaluation points in $\mathcal{X}$ and $\mathcal{Y}$.
This gives us a real-valued quantity $\Lambda$ which is defined as
\begin{multline}\label{eq:Lambda}
\Lambda := \sum_{i=0}^{n_0}\sum_{j=0}^{m_0} \alpha_i \beta_j f(x_i, y_j) \\
                = \sum_{n,m\geq 0} q(m|n) \sum_{i=0}^{n_0} \alpha_i p_n(x_i) \sum_{j=0}^{m_0} \beta_j r_m(y_j),
\end{multline}
where coefficients $\{\alpha_i\}_{i=0}^{n_0}$ and $\{\beta_j\}_{j=0}^{m_0}$ are real numbers.
Also, we write
\be
u_n := \sum_{i=0}^{n_0} \alpha_i p_n(x_i),\quad
v_m := \sum_{j=0}^{m_0} \beta_j r_m(y_j),
\ee to capture the summation over all the considered evaluation points.

Here, we want to get $\Lambda$ as close as possible to $q(m^*|n^*)$.
To do that, we set $u_n = \delta_{n,n^*}$ and $v_m = \delta_{m,m^*}$ for all values of $n \in \mathcal{N}_0$ and $m \in \mathcal{M}_0$, where $\delta_{a,b}$ is the Kronecker delta function, and solve for $\alpha_i$ and $\beta_j$.
In essence, this step requires solving two systems of linear equations, namely one for the source device,
\be \label{eq:solve_alphas}
\underbrace{\begin{bmatrix}
    p_0(x_0)       & p_0(x_1) & \hdots & p_0(x_{n_0}) \\
   p_1(x_0)       & p_1(x_1) & \hdots & p_1(x_{n_0}) \\
    \vdots       & \vdots & \ddots & \vdots \\
   p_{n_0}(x_0)       & p_{n_0}(x_1) & \hdots & p_{n_0}(x_{n_0})
\end{bmatrix}}_{\rm{Input~photon-number~distribution}}
\begin{bmatrix}
    \alpha_0  \\
    \alpha_1  \\
    \vdots  \\
    \alpha_{n_0}
\end{bmatrix}
=
\begin{bmatrix}
    \delta_{0,n^*}  \\
    \delta_{1,n^*}  \\
    \vdots  \\
    \delta_{n_0,n^*}
\end{bmatrix},
\ee and one for the measurement device,
\be\label{eq:solve_betas}
\underbrace{\begin{bmatrix}
    r_0(y_0) & \hdots & r_0(y_{m_0}) \\
   r_1(y_0) & \hdots & r_1(y_{m_0}) \\
    \vdots        & \ddots & \vdots \\
   r_{m_0}(y_0) & \hdots & r_{m_0}(y_{m_0})
\end{bmatrix}}_{\rm{Detection~response~function}}
\begin{bmatrix}
    \beta_0  \\
    \beta_1  \\
    \vdots  \\
    \beta_{m_0}
\end{bmatrix}
=
\begin{bmatrix}
    \delta_{0,m^*}  \\
    \delta_{1,m^*}  \\
    \vdots  \\
    \delta_{m_0,m^*}
\end{bmatrix}.
\ee Notice that this step does not require the knowledge of $f(x_i,y_j)$ and hence can be seen as part of the calibration process prior to characterising the channel.

Solving Eq.~\eqref{eq:solve_alphas} and Eq.~\eqref{eq:solve_betas} hence gives
\begin{multline}\label{eq:residual}
\Lambda = q(m^*|n^*)\\
            + \sum\limits_{n\geq n_0+1} q(m^*|n)u_n
            + \sum\limits_{m\geq m_0+1} q(m|n^*)v_m \\
            + \sum\limits_{n\geq n_0+1}\sum\limits_{m\geq m_0+1} q(m|n) u_n v_m.
\end{multline}
As one can see, $\Lambda$ is now expressed in terms of the desired quantity, $q(m^*|n^*)$, and some other irrelevant terms that emanate from higher photon number contributions, i.e. those from $n \geq n_0+1$ and $m \geq m_0+1$.
In Appendix \ref{app:bound_derivation}, we show that these terms can be rigorously bounded by using known information of the optical devices. This in turn provides upper and lower bounds on $q(m^*|n^*)$.
More concretely, the idea is to establish bounds on $u_n$ and $v_m$ using the characterised input photon-number distribution and detection response function.
Therefore, these bounds are specific to the types of light sources and detectors used in the setup; in Appendix \ref{app:bound_derivation}, we provide standard bounds for common optical devices such as phase-randomised lasers, threshold detectors and homodyne detectors with phase randomised local oscillators.
We also note that these bounds can be made arbitrarily tight by selecting large enough $n_0$ and $m_0$ values.
Indeed, a key condition is to ensure that the derived bounds on the extra terms in Eq.~\eqref{eq:residual} are small when compared to $q(m^*|n^*)$; and this can be achieved by using bigger values of $n_0$ and $m_0$.

This linear method is conceptually similar to early papers in homodyne tomography using \emph{pattern functions} to recompute photon-number statistics~\cite{munroe_photon-number_1995, leonhardt_sampling_1996}.
They considered the use of pattern functions $M_n(x)$ such that:
\begin{equation}
  \label{eq:pattern-fctn}
  p_n = \int_{-\infty}^{+\infty} M_n(x)f(x) dx.
\end{equation} Indeed, there is an obvious similarity with our method when using only one input mode:
\begin{equation}
  \label{eq:pattern-fctn-discretised}
  \Lambda = \sum_{i=0}^{n_0} \alpha_i f(x_i),
\end{equation}
and $\Lambda$ is a good approximation of $p_n$ up to some deviation we can bound.
Our computation in Eq.~\eqref{eq:pattern-fctn-discretised} can be seen as a discretized version of Eq.~\eqref{eq:pattern-fctn}.
The main benefit is that it can be generalised easily beyond homodyne tomography, for instance using threshold detectors or other light sources, and as such may provide a unified understanding of photon-number probability estimation.

\section{Applications}
\label{sec:applications}
We present here two applications to illustrate the utility of our framework.
In the first application, we consider the security of practical prepare-and-measure QKD with realistic photon sources and single photon detectors.
More specifically, we show how to rigorously bound the single-photon channel security of the protocol.
In the second application, we show how to use our framework to enable TCSPC with homodyne detection instead of single photon detection.

\subsection{Prepare-and-measure QKD \\ with single-photon channel security}

As a first application of our method, we analyse the asymptotic security analysis of discrete-variable QKD protocols based on practical optical devices such as lasers and threshold detectors.
Here, we consider the celebrated Bennett-Brassard 1984 (BB84) protocol \cite{bennett_quantum_2014} and the six-state protocol \cite{brus_optimal_1998,bechmann-pasquinucci_incoherent_1999}.
These two protocols are formulated as \textit{qubit} protocols, i.e. the preparation, evolution and measurement can be described in a Hilbert state of dimension 2 \cite{scarani_security_2009}.

In practice, however, most QKD systems use weak coherent laser sources and threshold detectors~\cite{scarani_security_2009} to implement qubit states and measurements.
This is because, as mentioned above, deterministic single-photon sources and PNRDs are not yet available and weak coherent laser sources and threshold detectors are the closest one can get to achieving qubit states and measurements in practice (at least with regards to cost and practicality).
However, one critical drawback is that there are some fundamental differences between the qubit models (assumed in the protocol) and these optical devices.
Some of these differences are irreconcilable and hence cannot be applied to qubit protocols, while some may lead to implementation loopholes such as photon-number splitting attacks~\cite{huttner_quantum_1995,lutkenhaus_quantum_2002}.

Fortunately, for most qubit protocols (including BB84), the gap between theory and practice can be mitigated using innovative techniques such as the decoy-state method~\cite{hwang_quantum_2003, lo_decoy_2005, wang_beating_2005} and squashing \cite{gottesman_security_2004, beaudry_squashing_2008,tsurumaru_squash_2010, fung_universal_2011}.
In the case of the former, assuming that the prepared optical signals are diagonal in the photon-number basis, Eve's attacks can be categorised according to the emitted photon numbers.
This allows us to focus on the single-photon component of emitted optical signal and hence view the states prepared by Alice as qubits.
On the other hand, squashing models provide an elegant method to map the actual full (infinite dimensional) mode measurement onto a finite dimensional Hilbert space followed by the ideal measurement.
In the case of the BB84 protocol, a squashing model exists and hence one could assume qubit models for Bob's measurements in practice.

Therefore, using the decoy-state method together with a squashing model, one can derive statistical bounds on the single-photon error rates of the BB84 protocol and hence compute its secret key rate.
However, this approach is quite restrictive and does not apply readily to other qubit protocols. For instance, it has  been shown that squashing does not immediately apply to the six-state protocol~\cite{beaudry_squashing_2008,tsurumaru_security_2008}; on the other hand, it has been shown that by relaxing certain statistical constraints, it is possible to define squashing maps for a wide range of finite-dimensional protocols \cite{fung_universal_2011}.

By contrast, our method allows us to analyse any qubit (more generally, any higher dimension) protocol~\cite{cerf_security_2002,islam_securing_2018,islam_scalable_2019} without using a squashing model.
Assuming that Alice and Bob prepares and receives a single photon defined across two orthogonal optical modes, the states and measurements can be described by qubit states and qubit measurements, respectively.
Our method allows us to bound the probability of Bob receiving a single photon (just before the measurement) when Alice prepares and sends a single photon as well as the corresponding error rate in each basis.

Roughly speaking, our method provides three practical advantages over existing methods.
Firstly, since we consider only secret key contributions from events in which Alice prepares a single photon and Bob receives a single photon, we can directly use any security proof technique for qubit models without applying any squashing model.
As such, our method can be applied to most practical QKD systems under the condition that these systems randomly vary their detection efficiency as specified by the detector-decoy method.
Secondly, since squashing models typically require mapping double-detection events to random outcomes~\cite{beaudry_squashing_2008,tsurumaru_security_2008}, applying a squashing model would likely introduce some additional errors from the detector background noise.
On top of that, squashing also does not differentiate between clicks due to true single photon detections and empty detections, which may also introduce additional errors.
Hence, our method, which can rigorously bound the true channel error rates, could give an enhanced bound on the secret key rate especially in the high loss regime where the dark count rate is not negligible.
Finally, as mentioned, our method allows us to analyse the security of the protocol based on single-photon channel security.
To appreciate this feature better, we comparatively note that when using the decoy state method combined with a squashing model, one actually evaluates the security of the channel together with the detector noise, which is normally trusted.
In this case, the single-photon error rates includes the trusted detector noise.
By contrast, our method allows the rigorous separation of channel noise and detector noise and thus provides a concise method to derive lower bounds on the secret key rate in the calibrated detector setting; in fact, the security of QKD with calibrated devices is known to be an open problem~\cite{scarani_security_2009} .

We now demonstrate how our method can be applied to the security analyses of practical QKD systems. We emphasise that our method can be used for most discrete-variable protocols, but as concrete examples, we only apply our method to BB84 and six-state protocol.
To that end, we introduce some notations that we use in this subsection. We denote the basis choice of Alice and Bob by $x$ and $y$ respectively. $x$ and $y$ are randomly chosen from the set $\mathcal{X}$ and $\mathcal{Y}$ respectively where $\mathcal{X} = \mathcal{Y} = \{ X, Z\}$ for BB84 and $\mathcal{X} = \mathcal{Y} = \{ X, Y, Z\}$ for the six-state protocol.
The symbol value encoded by Alice is denoted by $a \in \mathcal{A}$ and Bob's detection pattern is denoted by $b \in \mathcal{B}$.
For both the BB84 and six-state protocol, $a \in \{0,1\}$ and $b = b_0 b_1$ is a two-bit string where $b_i$ indicates whether the detector in mode $i$ clicks (we have $b_i = 0$ when the detector in mode $i$ does not click and $b_i = 1$ when the detector clicks).
Based on $b$, Bob would then map the observed click pattern into the decoded symbol or he could choose to discard inconclusive events (such as no-click or double-click events).
Finally, we denote Alice's intensity setting by $\mu$ and Bob's detection efficiency by $\eta$ (composed of $\eta_0$ for detector $0$ and $\eta_1$ for detector $1$) which are chosen randomly in each round.
We will now focus our attention on the case where $\abs{\mathcal{A}} = 2$, i.e., the protocol uses binary symbols. The generalisation to a higher dimensional protocol is straightforward.

In a typical discrete-variable protocol, Alice would randomly choose a basis $x$ and bit value $a$.
She would then choose an intensity setting $\mu$ and then prepare a phase-randomised coherent state in the corresponding mode.
For phase-randomised coherent source, the emitted photon number $n$ would follow a Poisson distribution with mean $\mu$.
Similarly, Bob randomly chooses a basis choice $y$ as well as detection efficiency setting $\eta$ and he obtains the outcome $b$.
In the parameter estimation step, Alice and Bob can estimate the following conditional probabilities (for all possible combination of parameters):
\be
f^{\{xyab\}} (\mu, \eta) = \Pr(b | x,y,a, \mu, \eta).
\ee
This function can be expanded as
\begin{multline}
\label{eq:f_qkd}
f^{\{xyab\}} (\mu, \eta) \\= \sum\limits_{n\geq 0} p_n(\mu) \underbrace{\sum_{k,l\geq 0}q^{\{xya\}}(kl |n) r_k^{\{b_0\}}(\eta_0) r_l^{\{b_1\}}(\eta_1)}_{Y_n^{\{xyab\}}},
\end{multline}
where $p_n(\mu)$ is the probability of the source emitting $n$ photons (defined in Eq.~\eqref{eq:laser}) while $q^{\{xya\}}(kl |n)$ is the probability of $k$ photons arriving in mode 0 and $l$ photons arriving in mode 1 given that the source emitted $n$ photons, Alice chooses the basis $x$ and bit value $a$ and Bob chooses measurement basis $y$.
$r_k^{\{b_0\}}(\eta_0)$ denotes the probability of the detector in mode 0 clicking ($b_0=1$) or not ($b_0=0$), given that $k$ photons arrived in that mode and $r_l^{\{b_1\}}(\eta_1)$ is defined similarly. Note that the probability of a threshold detector not clicking is given in Eq.~\eqref{eq:threshold}.

As such, Eq.~\eqref{eq:f_qkd} represents the conditional probability as a product of different system elements (transmitter, channel and receiver) in the form given in Eq.~\eqref{eq:meas_dist}.
The parameters $x$, $y$, $a$ and $b$ are considered as fixed parameters when estimating the input-output photon-number distribution. Here, we remark that these parameters serve only as a means for Alice and Bob to organise their measurement data, i.e. they categorise the input-output photon-number distributions according to $x$, $y$, $a$ and $b$. Of note, the input-output photon-number distributions have to be independent of these parameters: this is needed to ensure that the single-photon channel behaviour is \emph{basis-independent}. In other words, Eve's attacks on the quantum channel have to be independent of Alice's and Bob's basis choices~\cite{gottesman_security_2004}. In addition to this, we also need that the input-output photon-number distributions are independent of $\mu$ and $\eta$. In practice, these conditions can be reasonably enforced by using phase-randomised coherent lasers and photon-counting detectors, which is the case in our consideration.

Notice that the $n$-photon yield, which is denoted by $Y_n^{\{xyab\}}$ here, is the usual quantity of interest in decoy-state QKD \cite{lo_decoy_2005}.
More precisely, in Ref.~\cite{lo_decoy_2005}, the authors considered events in which Alice and Bob choose the same basis, i.e. $x=y$, and the cases in which Bob observes at least one click.
They also average the yield over Alice's bit value $a$.
Hence, the $n$-photon yield is the probability of observing a click given that Alice's laser emits $n$ photons.
Clearly this would depend on both the channel and the trusted detectors which are located in Bob's lab.
In contrast, our method bounds the probability of $k$ and $l$ photons arriving at Bob's measurement device given that $n$ photons are prepared by Alice.
This would depend only on the behaviour of the channel and not on the detectors.

\begin{figure}[t!]
        \includegraphics[width=\columnwidth]{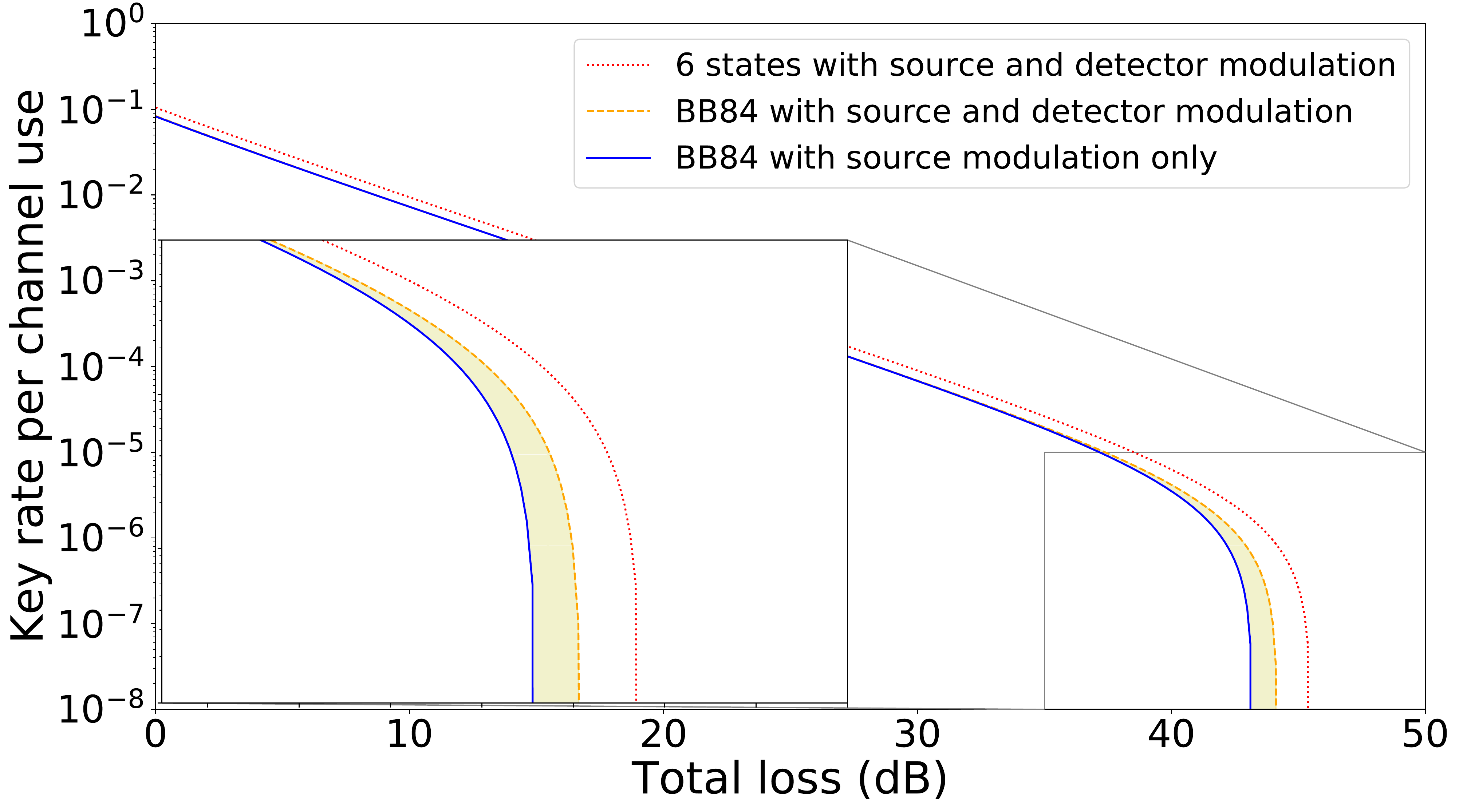}
        \caption{We simulate the achievable secure key rate with two avalanche photo-diode detectors, each one featuring a dark count rate of $10^{-6}$ and a fixed channel error rate of $5\%$ in all bases (depolarising channel).
            The secure key rate in the source and detector modulation case is $K \geq p_0(\mu)q_{00|0}\big(1- r_0^{(0)}(\eta_0)r_0^{(0)}(\eta_1)\big) + p_1(\mu)p^{x=y=0}_\text{det} H(A|E) - Q(\mu, \eta)h_2\big(E(\mu,\eta) \big)$ where $H(A|E)$ is the conditional entropy on Alice's key bit given Eve's side information for a qubit protocol. For BB84, we have $H(A|E)\geq 1-h_2(e_X)$, for six states we have $H(A|E) \geq 1 - \Big( H(\lambda) - h_2(e_Z) \Big)$. To recompute the single photon statistics, we use three intensity levels: $10^{-3}, 10^{-2}, 0.5$ and four efficiency levels per detector: $0.94, 0.96, 0.98, 1$. See more details in Appendix~\ref{app:qkd}.}
            \label{fig:qkd_plot-1}
    \end{figure}

    \begin{figure}[t!]
        \includegraphics[width=\columnwidth]{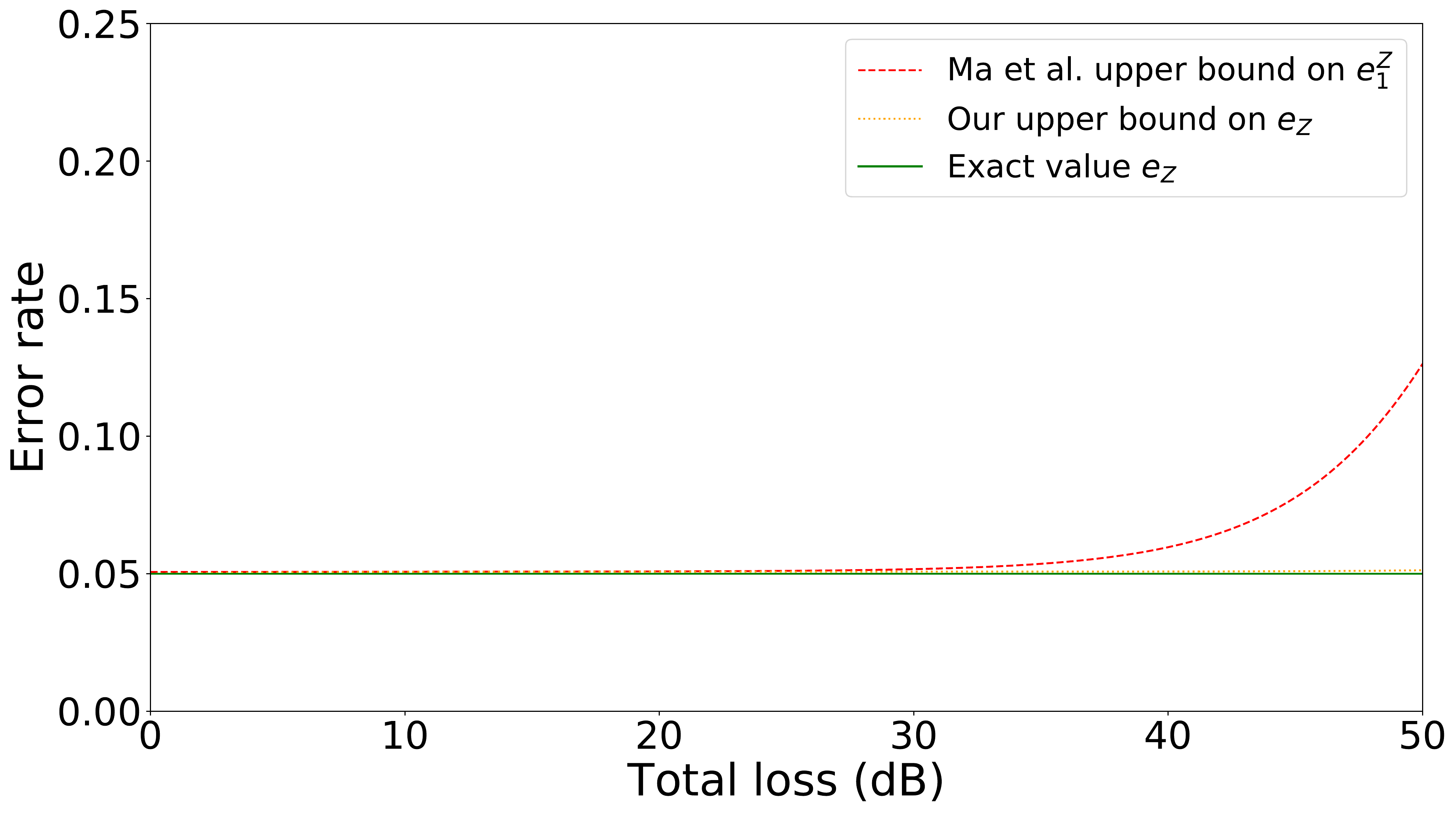}
        \caption{We compare the exact channel error rate value used for the simulation ($5\%$ here) to the upper bound provided by our proposed analytical method and the one proposed in Ref.~\cite{ma_practical_2005}. The definition in Ref.~\cite{ma_practical_2005} is including the noise from the detector dark counts while our is considering only the noise on the channel. As a result, the definition of Ref.~\cite{ma_practical_2005} is increasing while ours is staying close to the exact value in the high loss regime, hence a slight improvement in distance for the key rate. Note that this improvement is only affecting privacy amplification, since the noise due to dark counts still has to be corrected in the error correction step.}
        \label{fig:qkd_plot-2}
\end{figure}

To analyse the security of the BB84 and six-state protocol, the quantity of interest is $q^{\{xya\}}(kl | n)$ when $n=1$ and $k+l = 1$, i.e. the probability of the channel outputting a single photon given that a single photon enters the channel.
In this case, the conventional security proofs \cite{shor_simple_2000,renner_information-theoretic_2005} that rely on the qubit models can be directly applied without the need for squashing model \cite{gottesman_security_2004, beaudry_squashing_2008, fung_universal_2011}.
For all values of $x, y, a$, it is possible to use either method of Section \ref{sec:methods} to deduce $\{q^{\{xya\}}(10 | 1),q^{\{xya\}}(01 | 1)\}$ once $f^{\{xyab\}} (\mu, \eta)$ is obtained from the parameter estimation step of the protocol.
The single-photon error rates are just functions of $\{q^{\{xya\}}(10 | 1),q^{\{xya\}}(01 | 1)\}$; for example, the single-photon error rate given that Alice and Bob have chosen the same basis (i.e. $x = y$) is given by
\begin{multline}
    e_{x=y}
    \\= \frac{\sum\limits_{a} \Pr(a) q^{\{xya\}}(a,a \oplus 1|1) \left( 1 - r_{a}^{\{0\}} (\eta_0) r_{a\oplus 1}^{\{0\}} (\eta_1) \right)}{p_\text{det}^{x=y}}
\end{multline}
where the qubit detection probability is
\begin{multline}
    p^{x=y}_\text{det} \\=
    \sum\limits_{a}\sum\limits_{k+l=1} \Pr(a) q^{\{xya\}}(k,l |1) \left(1 - r_k^{\{0\}}(\eta_0)r_l^{\{0\}}(\eta_1) \right)
\end{multline}

 Once the single-photon error rates are determined, the secret key rate can be easily computed.
 Here, we present the bound on the secret key rate while we defer the detailed security analysis of the six-state protocol to Appendix \ref{app:qkd}.
 In the asymptotic limit and under the assumption of collective attacks, the secret key rate $K$ of BB84 and six-state protocols is given by
\begin{multline}\label{eq:keyrate}
K \geq  p_0(\mu)q(00|0)\big(1- r_0^{\{0\}}(\eta_0)r_0^{\{0\}}(\eta_1)\big) \\
 + p_1(\mu) p^{x=y=Z}_\text{det} H(A|E) \\
 - Q(\mu, \eta)h_2\left(E(\mu,\eta) \right),
\end{multline}
where $H(A|E)$ is single-photon conditional entropy given Eve's quantum side information; $h_2(\cdot)$ denotes the binary entropy function.
$Q(\mu, \eta)$ and $E(\mu,\eta)$ is the observed gain and quantum bit error rate when Alice and Bob choose intensity $\mu$ and detection efficiency $\eta$ respectively.
Hence, the first term is the contribution due to the events in which Alice prepares vacuum state.
Since no quantum information is leaked whenever Alice prepares the vacuum state, all the detected events due to the transmission of vacuum states are secure.
The second term is the single-photon contribution and the third term is the leakage due to error correction assuming that the error-correcting code saturates the Shannon limit.
Using $\{e_X, e_Y, e_Z\}$ as short-hand for the single-photon error rate in the respective basis, the single-photon conditional entropy $H(A|E)$ for the BB84 protocol and the six-state protocol is given by
\begin{equation}
    H(A|E) =
    \begin{cases}
        1 - h_2(e_X)  & \text{(BB84)}\\
        1 + h_2(e_Z) - H(\lambda) & \text{(six-state)}
    \end{cases}
\end{equation}
Here, $\lambda = (\lambda_0, \lambda_1, \lambda_2, \lambda_3)$ is a real vector containing the unique solution to the following simultaneous equations
\begin{equation}
\begin{matrix}
\lambda_0 &+& \lambda_1 &&&& &=& 1-e_Z\\
\lambda_0 &&&+&\lambda_2&& &=& 1-e_X\\
\lambda_0 &&&&&+&\lambda_3&=& 1-e_Y\\
\lambda_0 &+& \lambda_1 &+& \lambda_2 &+& \lambda_3 &=& 1\\
\end{matrix}
\end{equation}
and $H(\lambda)$ is the corresponding Shannon entropy. Therefore, by substituting the appropriate $H(A|E)$ to Eq.\eqref{eq:keyrate}, we obtain the bound on the secret key rate of the corresponding protocol.

We present the simulated secret key rates in Fig.~\ref{fig:qkd_plot-1} assuming standard SPADs parameters. Here, we compare against the asymptotic secret rate based on the standard decoy-state method~\cite{ma_practical_2005}.
As mentioned above, the key difference is that our method directly evaluates the single-photon channel security whereas the standard decoy-state method would include the detectors' background noise (dark counts).
Indeed, in Fig.~\ref{fig:qkd_plot-2}, we see that the single-photon error rate of
our method does not include the detectors' dark counts in the channel error rate and hence our proposed upper bound on the error rate remains close to the exact value while that based on Ref.~\cite{ma_practical_2005} is dominated by the dark count noise in the high loss regime.

\subsection{Time-Correlated Single Photon Counting with Homodyne detection}
\begin{figure}
        \includegraphics[width=6cm]{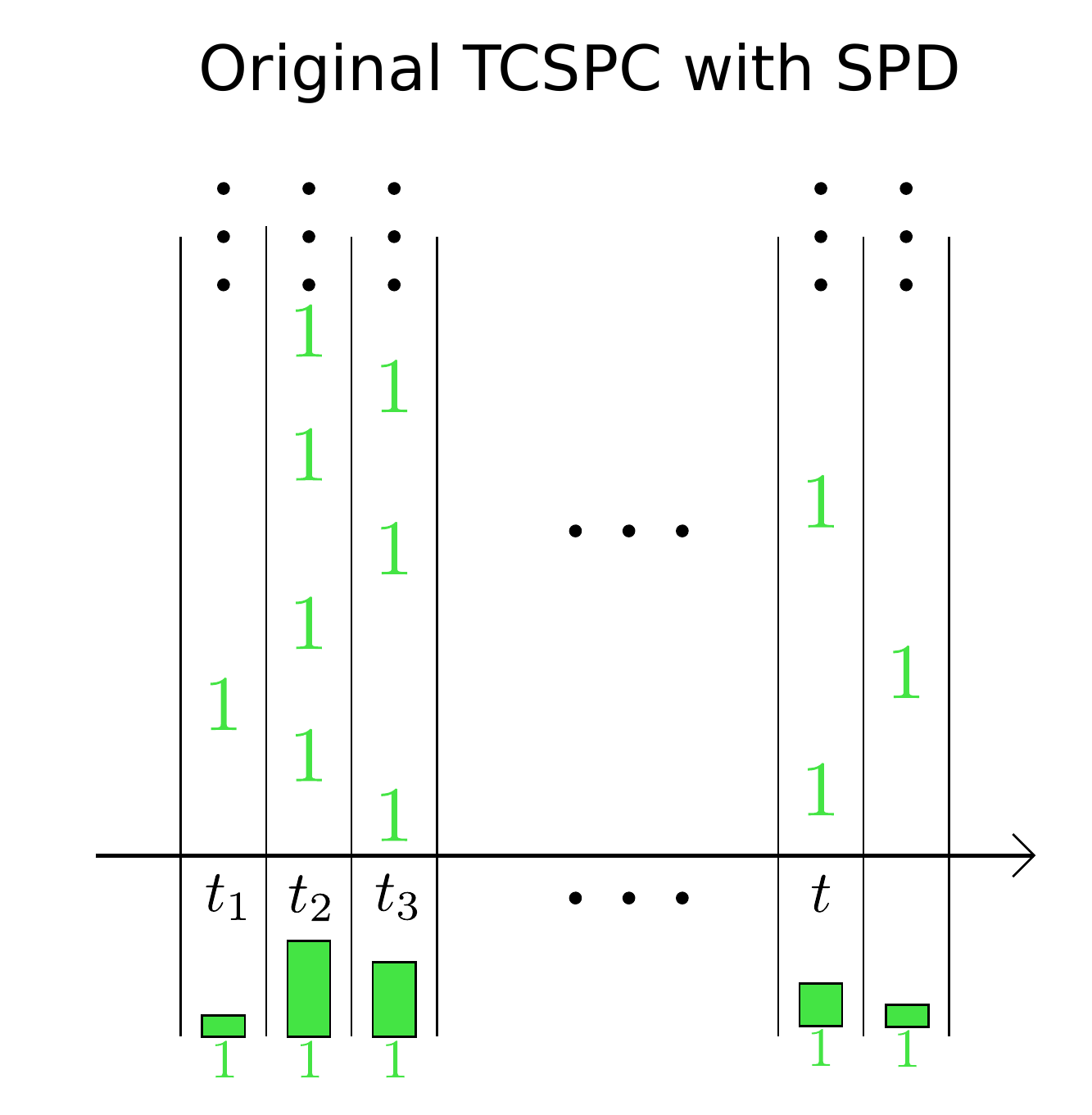}
        \caption{The original TCSPC concept is based on the recording of single-photon detection events and construction of the corresponding histogram as shown above. As a result, there is an implicit post-selection of the conclusive outcomes.
That is, the experiment is repeated until the bin with the maximum number of events reaches a certain level.
The histogram is then normalised to derive the probability of having a detection event in a particular bin $t \in \mathcal{T}$ given that there was a conclusive outcome $C$, i.e. $\Pr(t | C)$.
Nevertheless, due to the dead time effect, the recorded probability is not exactly the one aforementioned, but rather the probability of observing a detection in a bin and not recording any detection before. When the counting rate is low, the two probabilities are close though \cite{yguerabide_nanosecond_1972}. }
        \label{fig:tcspc-1}
\end{figure}

Here we propose TCSPC with homodyne detection instead of single photon detection.
There are significant practical benefits in doing so; largely one could reduce the implementation cost and footprint of TCSPC through integrated photonics platforms.
The basic idea of TCSPC is to measure the single photon emission time profile in the nanosecond time scale, e.g. fluorescence decays of excited samples \cite{yguerabide_nanosecond_1972}.
The current method achieves this via the small time resolution (hundreds of picoseconds) of Single Photon Detectors (SPDs; e.g. photo-multiplier tubes, micro-channel plates, SPADs) to measure the time difference between a reference ``start'' signal and a ``stop'' signal triggered by a single photon emission event \cite{becker_overview_2005, oconnor_basic_1984}.
As such, by using a pulsed laser the intensity profile can be sampled repeatedly and a histogram of photon arrivals per time bin over the intended time domain can be constructed; assuming the probability of multi-photon emission is negligible. This concept is depicted in Fig.~\ref{fig:tcspc-1}.

However, SPDs typically suffer from finite recovery time (or dead-time); consequently, the detector becomes inactive for a period of time after a first detection event (hundreds of nanoseconds to dozens of microseconds \cite{hadfield_single-photon_2009, eisaman_invited_2011}).
As such, any optical signal arriving during this time window will not be detected and this problem tends to bias the measurement results towards earlier detection events.
This is a well known issue called \emph{pulse pile-up}~\cite{phillips_time_1985}.
In practice, to mitigate this problem, a popular approach is to keep the multi-photon emissions low and the counting rate below 2\%--5\% or lower~\cite{phillips_time_1985}, e.g. by restricting the excitation power.
In this case, nothing is detected most of the time and once in a while a unique photon is detected and recorded.
This common approach of limiting the excitation power solves the multi-photon emission issue but leads to a longer acquisition time.

Yet, TCSPC is not making full use of the single-shot information of a photon being detected or not at a particular time window; it extracts only the average count rate for each time window.
This suggests that other forms of detection technology could be used instead, for instance, homodyne detection. Indeed, this possibility has already been discussed in Ref.~\cite{arecchi_photocount_1966}: the authors therein described a linear method to compute the moments of an unknown probability distribution using the moments of the outcome function obtained experimentally. Our proposal with homodyne detection essentially follows this idea: that we can recover the photon-number statistics with phase-randomised homodyne detection. To help fix ideas, in the following we first briefly describe an ideal version based on PNRDs.
We then provide a proof of concept simulation of the homodyne TCSPC technique.

\subsubsection{TCSPC with perfect photon-number-resolving detection}
\label{sssec:virtual_TCSPC}

\begin{figure}[t]
        \includegraphics[width=7cm]{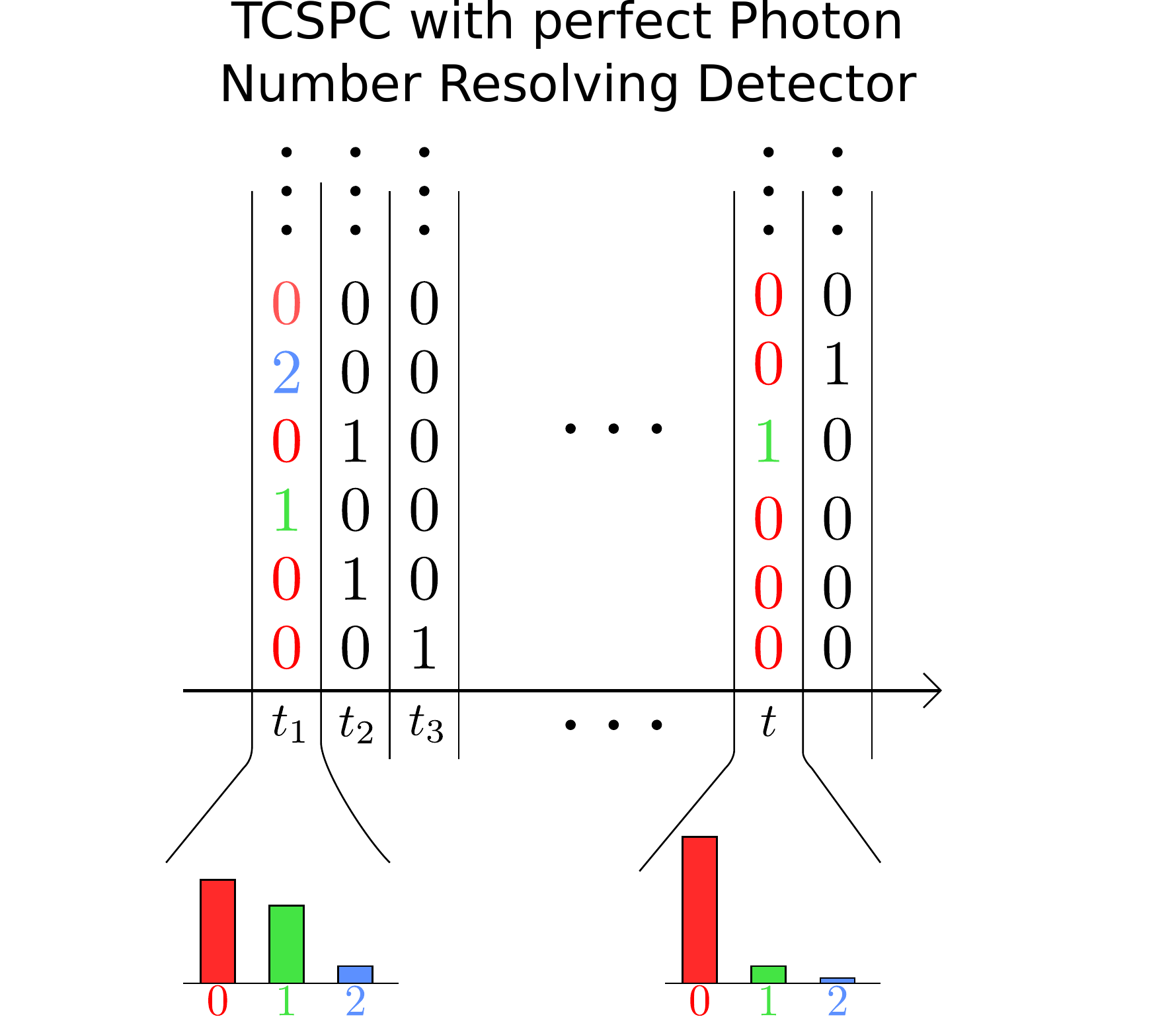}
        \caption{We consider here an ideal version of TCSPC where perfect PNRDs  are used to record the exact number of incoming photons in each time bin. We label $Y^{\{t\}} \in \mathbb{N}$ the corresponding random variable for each time bin. After enough events have been recorded, it is possible to estimate the photon number probability distribution at each time bin. By keeping only the single photon events, it is easy to recover the same TCSPC information as in Fig.~\ref{fig:tcspc-1}.
        }
        \label{fig:tcspc-2}
    \end{figure}

It is useful to first consider an intermediate ideal TCSPC protocol to illustrate the main ideas of our homodyne-based protocol. Here, we assume a perfect PNRD is used to measure the photons arrival time. By perfect, we mean that the detector has zero dead-time, perfect detection efficiency, and able to tell how many photons are detected in a given time window. Mathematically, the outcome of the protocol is described by a sequence of time-ordered random variables, $Y^{\{t\}} \in \mathbb{N}$, where each random variable counts the number of photons in the time bin $t \in T$ as shown in Fig.~\ref{fig:tcspc-2}.
Evidently, this ideal TCSPC protocol can recover the original TCSPC protocol's information by keeping only the events in which $Y^{\{t\}} \geq 1$.
We also highlight that there will be the same number of events recorded (regardless of the value of $Y^{\{t\}}$) in each time bin.
As a result, the probability of recording an event in a certain time window is $\Pr(t)=1/\abs{\mathcal{T}}$.

In the limit of many repetitions, the data from $Y^{\{t\}}$ allows computation the probability of detection of $n$ photons in the time bin $t$:
\begin{equation}
\label{eq:qn_t}
q_n^{\{t\}} = \Pr(Y^{\{t\}}=n) 
\end{equation}
We label $C$ a conclusive event ; it is the set of photon number values leading to a conclusive outcome, here all values $n \geq 1$.
We further denote the probability of a conclusive event within a time window:
\begin{equation}
q^{\{t\}} = \Pr(C|t)  = \sum\limits_{n\in C}q_n^{\{t\}}
\end{equation}
From this information only, it is possible to recompute the same probability as in the original version of TCSPC using the uniformity of $T$ and Bayes' rule:
\begin{equation}
\label{eq:bayes}
\Pr(t | C) = \frac{\Pr(C|t)\Pr(t)}{\sum\limits_{t'}\Pr(C|t')\Pr(t')} = \frac{q^{\{t\}}}{\sum\limits_{t'}q^{\{t'\}}}
\end{equation}

\subsubsection{Homodyne TCSPC}
\label{sssec:homodyne_TCSPC}
\begin{figure}[t]
        \includegraphics[width=7cm]{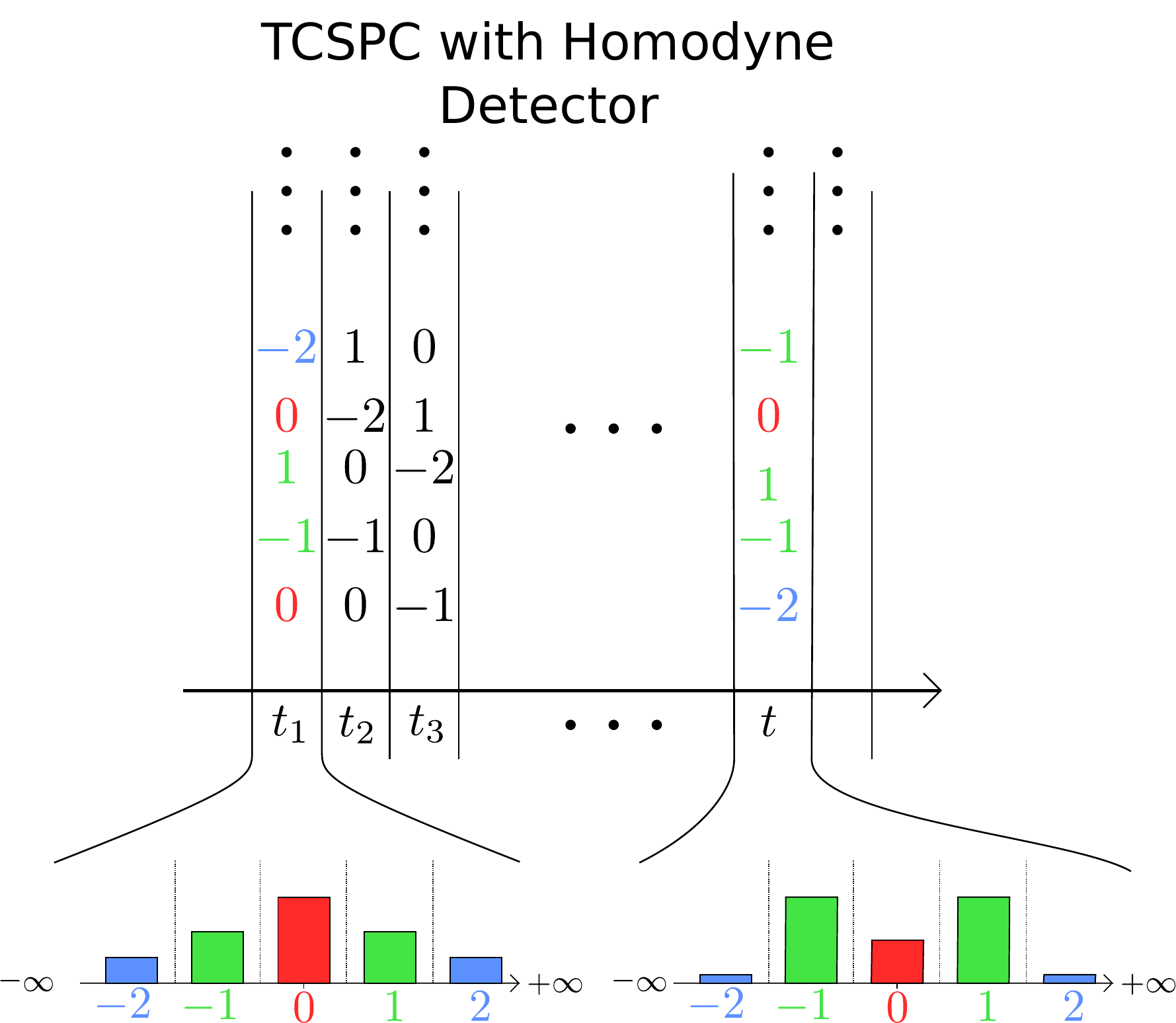}
        \caption{We consider here a practical implementation of TCSPC using a binned homodyne detector. We label $X^{\{t\}} \in \mathcal{X}$ the corresponding random variable for each time bin. After enough events have been recorded, it is possible to estimate the measurement PDF at each time bin. Then from this information, we show that it is possible to recompute the same photon number distribution as in Fig.~\ref{fig:tcspc-2}.
        }
        \label{fig:tcspc-3}
    \end{figure}

We now replace the perfect PNRD in the ideal version described above by a homodyne detector that sequentially measures all the time bins one after the other, up to the time resolution and speed of the ADC. In this case, the homodyne detector gives a continuous outcome which is also binned depending on the ADC resolution. We label $\mathcal{X}$ the set of bins.
Now the sequence of random variables $Y^{\{t\}} \in \mathbb{N}$ is replaced by $X^{\{t\}} \in \mathcal{X}$ as drawn in Fig.~\ref{fig:tcspc-3}.

In the limit of many repetitions and small bins in $\mathcal{X}$, we can recompute the probability density function (PDF) of every time bin which is assumed to have the following structure:
\begin{equation}
\label{eq:pdf_homodyne}
f^{\{t\}}(x) = \sum\limits_{n\geq 0} q_n^{\{t\}} A_n(x)
\end{equation}
with $A_n(x)$ representing the homodyne measurement response as we define in Section \ref{sec:models} Eq.~\eqref{eq:homodyne}.
It is also possible to consider a binned version for $f$ by using the binned version of $A_n(x)$ accordingly. Due to the structure of the detector, $f^{\{t\}}(x)$ is even i.e. $f^{\{t\}}(-x)=f^{\{t\}}(x)$, hence we can simply consider only the positive region, i.e. $x\geq 0$. The $q_n^{\{t\}}$ in Eq~\eqref{eq:pdf_homodyne} are defined in Eq~\eqref{eq:qn_t} and represent the contribution to the outcome due to $n$ photons on the detector. From the value of $f^{\{t\}}(x)$ at each $x \in \mathcal{X}$, the direct application of Section \ref{sec:methods} allows us to recompute a reasonably good estimation of the weights $q_n^{\{t\}}$ in front of the $A_n(x)$ in Eq~\eqref{eq:pdf_homodyne} for the low photon number events. Then recomputing the probability distribution as in the original TCSPC can be done with Eq~\eqref{eq:bayes}.

\begin{figure*}           \includegraphics[width=2\columnwidth]{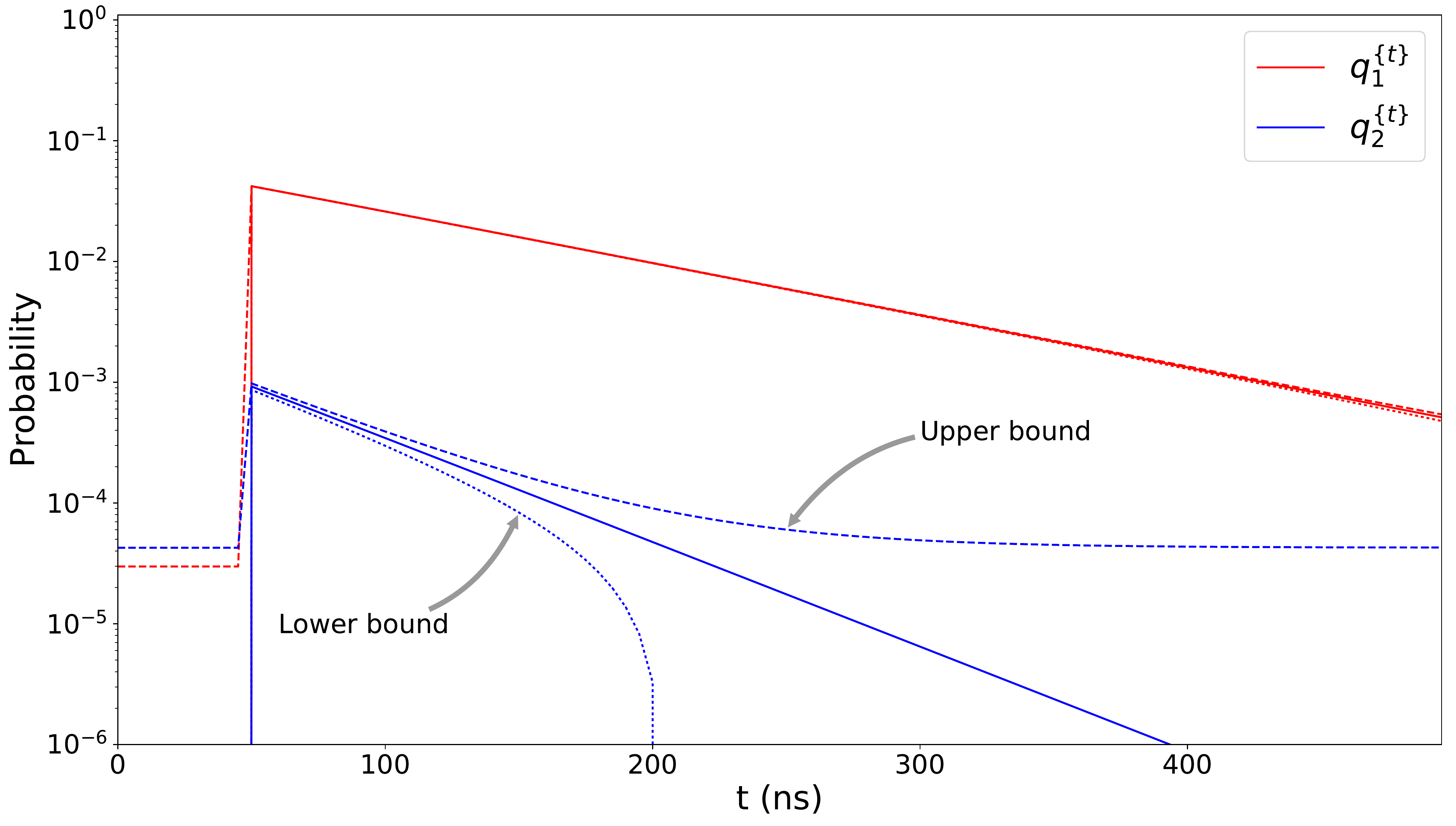}
            \caption{Simulation of an exponential decay intensity profile that can be reconstructed using homodyne detection only. The upper (dashed line) and lower (dotted line) bound on the probability of one photon emission are very close to the exact value (solid line) using our method while the bounds on the probability of two photon emission are looser due to the inability of the method to recover very low values. This simulation is using $16$ bins evenly spaced in the range $[0, 5]$ for the real outcome given by the homodyne detector. }
\label{fig:tcspc-plot}
\end{figure*}

We consider a simple physical experimental model to highlight the feasibility of our homodyne TCSPC. Here, we consider the intensity profile of a fluorescence decay after a delta excitation happening at $t_0 = \SI{50}{\ns}$. Following Ref. \cite{mandel_fluctuations_1959} and assuming an exponential decay for the intensity profile, the photon number distribution is
\be
q_n^{\{t\}} = \frac{\exp{\big(-E(t)\big)}E(t)^n}{n!},
\ee where $E(t)$ is the energy arriving on the detector for $t\geq t_0$
\begin{multline}
E(t)= \\ \int_t^{t+T} \alpha P(u) du = \alpha\exp{\Big(-\frac{t-t_0}{\tau}\Big)}\Big(1-\exp{\big(-\frac{T}{\tau}\big)\Big)}
\end{multline} and $T=\SI{5}{\ns}$ is the time bin duration, $\tau=\SI{100}{\ns}$ is the decay time, $\alpha=0.9$ is a coefficient including the excitation power and the detector sensitivity. Then, the measurement PDF is computed according to Eq.~\eqref{eq:pdf_homodyne} and a direct application of Section \ref{sec:methods} allows us to compute upper and lower bounds on the single and two photon emission probabilities.
Here, we use $16$ bins evenly spaced over the range $[0, 5]$ and apply the analytical method presented in Section \ref{sec:methods} to obtain the results shown in Fig.~\ref{fig:tcspc-plot}.
The bounds on the single photon emission probability are very close to the exact value for any time considered in our simulation while the bounds for the two photon emission probability are less tight due to its lower value that cannot be estimated well via our method.
This is due to the finite value of the extra terms in Eq.~\eqref{eq:residual} as discussed in Section \ref{sec:methods}.
One way to tighten this bound would be to consider more bins for the PDF.

\section{Conclusion}
In this paper, we show the possibility of using realistic light sources and detectors to recompute relevant information about the input-output photon number distribution of any unknown channel.
We describ a simple linear framework to model characterised sources, detectors and any multi-mode unknown channel.
Then, we present two computational methods to derive upper and lower bounds on the input-output photon number distribution.
Such information can be used for various applications in quantum optics and quantum information processing.
To that end, we highlight two applications: the single-photon channel security of practical QKD and TCSPC with homodyne detection.
For example, the application to QKD shows that this framework is a bridge between practical implementations and theoretical qubit security proofs.
Here, it is useful to mention that even though we present only one class of qubit protocols and an associated security proof, other finite-dimensional QKD protocols can also be proven secure in the same fashion, such as the Reference Frame Independent \cite{laing_reference-frame-independent_2010}, Loss-Tolerant \cite{tamaki_loss-tolerant_2014} and Tomography-based \cite{watanabe_tomography_2008, zhan_tomography-based_2020} protocols.
We can also highlight that the single photon detectors could possibly be replaced by homodyne detectors in certain schemes \cite{qi_bennett-brassard_2021}. The TCSPC example also suggests that this technology could possibly be deployed with homodyne detection for more cost-effective implementations.
Therefore, many applications relying on TCSPC as a module could potentially be upgraded to use a homodyne TCSPC instead. For instance, we can think of applications based on time-of-flight measurement \cite{massa_time--flight_1998} such as ranging \cite{ren_laser_2011, mccarthy_kilometer-range_2013} and low light imaging \cite{pawlikowska_single-photon_2017, tobin_three-dimensional_2019}.

\section*{Acknowledgements}
We acknowledge funding support from the National Research Foundation of Singapore (NRF) Fellowship Grant (No. NRFF11-2019-0001) and NRF Quantum Engineering Programme 1.0 Grant (No. QEP-P2) and the Centre for Quantum Technologies.

\medskip

\bibliography{main}

\onecolumngrid

\appendix

\section{Bound derivation}
\label{app:bound_derivation}
The goal of this section is to show that the estimate given in Eq.~\eqref{eq:residual} approximates well the quantity of interest $q(m^*|n^*)$. We give practical bounds on the extra terms. First we describe the general strategy and arguments that we use regardless of the actual hardware under use, and then we give the explicit bounds for a laser source, a threshold detector and a homodyne detector.

We recall a few notations:
\begin{align}
f(x,y) &= \sum\limits_{n,m\geq 0} p_n(x) q(m|n)r_m(y)\\
\Lambda &= \sum\limits_{i=0}^{n_0}\sum\limits_{j=0}^{m_0} \alpha_i \beta_j f(x_i, y_j) = \sum\limits_{n,m\geq 0} q(m|n) u_nv_m\\
u_n &= \sum\limits_i \alpha_i p_n(x_i)\\
v_m &= \sum\limits_j \beta_j r_m(y_j)
\end{align}

Let us define the residuals:
\begin{equation}
\label{def_residuals}
\begin{matrix*}[l]
R_{n_0} &=&  \sum\limits_{n\geq n_0+1} q(m^*|n) u_n \\
R_{m_0} &=& \sum\limits_{m\geq m_0+1} q(m|n^*)  v_m \\
R_{n_0m_0} &=& \sum\limits_{n \geq n_0 +1}\sum\limits_{m\geq m_0+1} q(m|n) u_n v_m
\end{matrix*}
\end{equation}

\subsection{General strategy}

The general strategy is to compute upper and lower bounds on these individual residuals to obtain bounds on $R = R_{n_0}+R_{m_0}+R_{n_0m_0} = \Lambda - q(m^*|n^*)$.
We highlight below a few general remarks that are useful for the subsequent derivation.

\begin{enumerate}
\item The sequences $u_n$ and $v_m$ depend respectively on $n_0$ and $m_0$, since it is the number of terms in the summation. The values of $\alpha$ and $\beta$ also depend on $n_0$ and $m_0$. Hence one has to be careful when considering $n_0 \text{ or } m_0 \to +\infty$
\item The series $u_n$ has a finite zeroth moment ($\sum\limits_{n\geq0} u_n < +\infty$) and first moment ($\sum\limits_{n\geq0} n u_n < +\infty$). That is because it is related to the probability distribution of a source with finite energy.
\item The series $v_m$ does not have a finite zeroth moment in general since it is related to a conditional probability. However, the sequence is always bounded.
\item $\forall n,m \geq 0: 0 \leq q(m|n) \leq 1$
\item The series $q(m|n)$ has a finite zeroth moment in $m$ for all $n$:  $\sum\limits_{m\geq0} q(m|n) = 1$
\item The series $q(m|n)$ does not have a finite zeroth moment in $n$ in general.
\end{enumerate}

\subsubsection{Bound on $R_{n_0}$}
The best we can do for bounding $R_{n_0}$ is to bound $q(m^*|n)$ by $+1$ or $-1$ depending on the sign of $u_n$.
Let us denote:
\begin{align}
u_n^{\oplus} &= \max(0, u_n) \geq 0\\
u_n^{\ominus} &= \min(0, u_n) \leq 0
\end{align}

Then the bound is related to the remainder of the convergent series $u_n^{\oplus}$ and $u_n^{\ominus}$:
\begin{equation}
\sum\limits_{n\geq n_0+1} u_n^{\ominus} \leq R_{n_0} \leq \sum\limits_{n\geq n_0+1} u_n^{\oplus}
\end{equation}

\subsubsection{Bound on $R_{m_0}$}
For this one, we need to combine the series $q(m|n)$ and $v_m$ together since the former is convergent in $m$ while the latter is not.
Similarly, let us denote:
\begin{align}
v_m^{\oplus} &= \max(0, v_m) \geq 0\\
v_m^{\ominus} &= \min(0, v_m) \leq 0
\end{align}
Additionally we assume that the sequence $v_m^{\oplus}$ is non-increasing for $m\geq m_0+1$ and similarly, $v_m^{\ominus}$ is non-decreasing for $m\geq m_0+1$.
If it is not the case, we assume that we can find an upper (lower) bound on $v_m^{\oplus}$ ($v_m^{\ominus}$) that satisfies this property, and we use it instead.
This assumption helps to compute an upper bound and lower bound on $v_m$ by simply considering the first element:
\begin{align}
\max\limits_{m\geq m_0+1}{v_m} &= v_{m_0+1}^{\oplus} \\
\min\limits_{m\geq m_0+1}{v_m} &= v_{m_0+1}^{\ominus}
\end{align}

In that case, we can derive the following bound:

\begin{equation}
v_{m_0+1}^{\ominus} \sum\limits_{m \geq m_0+1} q_{m|n^*}  \leq R_{m_0} \leq v_{m_0+1}^{\oplus} \sum\limits_{m \geq m_0+1} q_{m|n^*}
\end{equation}

We can see that this bound depends on our ability to find a good bound on:
\begin{equation}
\tilde{q} = \sum\limits_{m \geq m_0+1} q_{m|n^*} = 1 - \sum\limits_{m=0}^{m_0} q_{m|n^*}
\end{equation}

As a first approximation, we can use $\tilde{q} \leq 1$.
It is possible to start from this rough estimate to obtain a valid upper and lower bounds on a few $q(m|n^*)$ in the range $m\in \set{0 \dots m_0}$ and then use this information to update the bound on $\tilde{q}$.
After a few iterations of this procedure, the bounds on $R_{m_0}$ usually become of the same order of magnitude as those on $R_{n_0}$. Alternatively, it is also possible to analyse some geometric-arithmetic sequence to compute the limit after many iterations.

\subsubsection{Bound on $R_{n_0m_0}$}
This bound is easy to compute since we do not have much information anyway. The best we can do is to bound $\sum\limits_{m\geq m_0+1} q_{m|n}$ by $1$.
\begin{equation}
\abs{R_{n_0m_0}} \leq \sum\limits_{n\geq n_0+1} \abs{u_n} \norm{v}_{\infty} \leq \sum\limits_{n\geq n_0+1} \max(u_{n_0+1}^{\oplus}, -u_{n_0+1}^{\ominus}) \max(v_{m_0+1}^{\oplus}, -v_{m_0+1}^{\ominus})
\end{equation}

The reader can notice that the bound on $R_{n_0m_0}$ is roughly the product of the previous bounds on $R_{n_0}$ and $R_{m_0}$ hence it is small in practice, and the main contribution comes from the residuals related to source only $R_{n_0}$ and detector only $R_{m_0}$. Similar to the previous cases, it is also possible to consider quantities like $\big(u_n v_m \big)^{\oplus}$ to refine $R_{n_0m_0}$ but the improvement is limited.

\subsection{Phase randomised laser}
We consider here a Poisson distribution for the source $p_n(x) = e^{-x}\frac{x^n}{n!}$ , and $n_0 + 1$ evaluation points $0\leq x_0 < x_1 < \dots < x_{n_0}\leq x_\text{max}$ with $x_\text{max}\leq  n_0$.

We notice that $p_n(x)$ is non-increasing in $n$ when $n$ is larger than $x_\text{max}$.
It is useful to keep the sign in $\alpha$ otherwise the bound quickly becomes loose, therefore it is not satisfactory to simply bound $\alpha$ by its 1-norm or positive or negative part.
However, due to the simple expression of $p_n(x)$, it is easy to check that $u_n$ has a constant sign.
Indeed, we find that $u_n \underset{\tiny{n \to +\infty}}{\sim} \alpha_{n_0}p_n(x_{n_0})$ so for $n$ large enough, the sign of $u_n$ is given by the sign of $\alpha_{n_0}$.
For the first few $n$ before that property is satisfied, it is possible to numerically check a finite number of values to establish a correct upper and lower bound. For the rest, we can take

\begin{align}
u_n^{\oplus} &= u_n \text{ and } u_n^{\ominus} = 0, \text{ if }u_{n_0+1} \geq 0\\
u_n^{\oplus} &= 0 \text{ and } u_n^{\ominus} = u_n, \text{ if }u_{n_0+1} < 0
\end{align}

In our examples, the sign is always constant and depending on the parity of $n^* - n_0$, similar to what Ref. \cite{tsurumaru_exact_2008} reported for decoy states with their bounds $X_n$ and $Z_n$ alternatively being lower or upper bound.

\subsection{Threshold detectors}
We consider $r_m(y) = (1-y)^m$ and $m_0 + 1$ evaluation points $y_0 \dots y_{m_0}$.
We denote $\tilde{y}_j = 1-y_j$ and assume $0\leq \tilde{y}_0 < \dots < \tilde{y}_{m_0}\leq \tilde{y}_\text{max}$ with $\tilde{y}_\text{max}$ small enough.
In other words, the $y_j$ are in a range $[1-\tilde{y}_\text{max}; 1]$.

We first notice that $r_m(y)$ is non-increasing in $m$ for any $y$. We also have the following property:
\be
r_{m+1}(y) = (1-y)r_m(y)
\ee
from which we find that $v_m$ is decreasing if it is positive and increasing if it is negative. Similar to the laser case, we find $v_m \underset{\tiny{m \to +\infty}}{\sim} \beta_{m_0}r_m(y_{m_0})$ where $y_{m_0}$ is the lowest value for $y$. Therefore the sign of $v_m$ is given by the sign of $\beta_{m_0}$ for $m$ large enough, then by monotonicity we find that the sign is constant for all $m$.

Therefore similar to the laser case, we take:
\begin{align}
v_m^{\oplus} = v_m \text{ and } v_m^{\ominus} = 0 \text{ if } v_{m_0+1} \geq 0\\
v_m^{\oplus} = 0 \text{ and } v_m^{\ominus} = v_m \text{ if } v_{m_0+1} <0
\end{align}

\subsection{Homodyne detectors}
We use a homodyne detector at the receiver.
The detection function is as follows \cite{tan_inverse_1997, banaszek_maximum-likelihood_1998}:

\begin{equation}
r_m(y) = A_m(y) = \sum\limits_{k=0}^m {m \choose k} \eta^k \big( 1-\eta \big)^{m-k} \abs{a_k(y)}^2
\end{equation}
which is a binomial mixture of Hermite functions $a_m(y)$ defined by the following recursion \cite{tan_inverse_1997}:
\begin{align}
a_{-1}(y) &= 0\\
a_0(y) &= \pi^{-\frac{1}{4}} \exp\Big( -\frac{y^2}{2}\Big) \\
a_{m+1}(y) &= \Big( \frac{2}{m+1}\Big)^{\frac{1}{2}} y a_m(y) - \Big( \frac{m}{m+1}\Big)^{\frac{1}{2}} a_{m-1}(y)
\end{align}

Since the functions $A_m(y)$ are even, we can assume without loss of generality that the outcome $y$ is non-negative. We further assume that $y \in [0 ; y_\text{max}]$.
This is motivated by the finite range of the ADC that will restrict the maximum observable value for the outcome.

We recall a few useful properties of $a_m(y)$ \cite{szego_laguerre_1939, indritz_inequality_1961, thangavelu_hermite_1993}:
\begin{align}
\label{eq:prop_am1}
\sqrt{2m}a_m(y) &= y a_{m-1}(y) - a_{m-1}'(y)\\
\label{eq:prop_am2}
a_m'(y) &= y a_{m-1}(y) - \sqrt{2m}a_m(y)
\end{align}

We define two functions for $m$ large enough to ensure $2m - y_{\text{max}}^2 > 0$:
\begin{align}
g_m(y) &= a_m(y)^2 + \frac{a_m'(y)^2}{2m+1 - y^2}\\
h_m(y) &= a_m(y)^2 + \frac{a_m'(y)^2}{2m - y^2}
\end{align}
$g_m(y)$ was defined by Szego in Ref. \cite{szego_inequalities_1939} to prove Sonin's theorem for Hermite functions. More specifically for our purpose, $g_m(y)$ is non-decreasing in $y$.

We can show using Eq.~\eqref{eq:prop_am1} and \eqref{eq:prop_am2} that $h_m(y)$ also satisfies \cite{indritz_inequality_1961}:
\be
h_m(y) = a_{m-1}(y)^2 + \frac{a_{m-1}'(y)^2}{2m - y^2}
\ee
and then $g_{m+1}(y) \leq h_{m+1}(y) \leq g_m(y) \leq h_{m}(y)$ hence $g_m(y)$ is non-increasing in $m$. Evidently, we also have $a_m(y)^2 \leq g_m(y)$.

We define as usual:
\begin{equation}
v_m = \sum_{j=0}^{m_0} \beta_j A_m(y_j)
\end{equation}

On the interval of interest, the functions $A_m(y)$ are oscillating between $0$ and some local maxima.
More precisely, the plot of $\abs{a_m}$ has $m+1$ ``bumps'' before quickly decreasing to zero.
When the detector efficiency is lower than $1$, the oscillations are attenuated for the first few local extrema, and the last bump remains predominant.
As a result, the series $v_m$ is also oscillating somehow, and it is difficult to analyse.

To simplify the analysis, we use an upper bound defined as follows for $m$ large enough to ensure $2m+1-y_\text{max}^2 > 0$:
\be
G_m(y) = \sum\limits_{k=0}^m {m \choose k} \eta^k \big( 1-\eta \big)^{m-k} g_k(y)
\ee

Then it is sufficient to take:

\begin{align}
v_m^{\oplus} &= \sum\limits_{j=0}^{m_)}\max(0, \beta_j) G_m(y_j)\\
v_m^{\ominus} &= \sum\limits_{j=0}^{m_)}\min(0, \beta_j) G_m(y_j)
\end{align}

\section{Single photon channel security of QKD}
\label{app:qkd}
We consider a binary six-state protocol where the key basis is given by the $\Z$-basis (also denoted basis $0$) and the test bases are given by the $\X$-basis and $\Y$-basis (denoted respectively $1$ and $2$).
The protocol operates as follows:
\begin{enumerate}
    \item \textbf{Preparation}: Alice randomly chooses a bit value $a \in \mathcal{A}=\set{0,1}$ uniformly \big($\Pr(a=0)=\Pr(a=1)=\frac{1}{2}$ \big) and basis $x \in \mathcal{X}=\set{0,1,2}$ with probability $\Pr(x)$.
    She also randomly selects the intensity $\mu \in M$ with probability $\Pr(\mu)$.
    The set $M$ has three intensity levels, and we assume that only the rounds where the highest intensity is chosen can be used to generate raw key bits. The other rounds are used for channel estimation only. In the simulation, we use $M=\set{10^{-3}, 10^{-2}, 0.5}$.
    She would then prepare phase-randomised coherent state with the chosen intensity $\mu$ to imprint her bit choice $a$ in basis $x$ on all photons in the pulse.

    \item \textbf{Measurement}: Bob draws two transmissivity values $\eta_0, \eta_1 \in E$ with probability $\Pr(\eta_0)$ and $\Pr(\eta_1)$ respectively.
    We write in short $\eta = (\eta_0, \eta_1)$ and their indices $j = (j_0, j_1)$.
    The set $E$ has four levels, and we assume that only the rounds using the highest value for both detectors are used to generate raw key bits. In the simulation we use $E=\set{0.94, 0.96, 0.98, 1}$.
    Bob also chooses a basis $y \in \mathcal{Y}=\set{0, 1, 2}$ with probability $\Pr(y)$.
    He will then set his variable attenuation for the first (second) detector to be $\eta_0$ ($\eta_1$) and perform a measurement in basis $y$.
    Finally, he records the measurement outcomes $b=b_0b_1 \in \mathcal{B}=\set{00,01,10,11}$.

    \item \textbf{Sifting}: After repeating step 1 and 2 for $N$ times, Alice and Bob use an authenticated classical channel to announce their basis choices, $x$ and $y$, and their modulation settings $\mu, \eta$.
    Then, for each tuple $(\mu_i, \eta_j)$, they partition the rounds into subsets $\Sset_{x,y,\mu_i,\eta_j}$ according to their basis, intensity and transitivity modulation choices.

    \item \textbf{Parameter estimation}: For rounds suitable for key generation, i.e. when $\mu=\max(M)$, $\eta_0=\eta_1=\max(E)$ and $x=y=0$, Bob will disclose a ---small--- random subset of his measurement outcomes $b$ when $b\neq 00$.
    For all other rounds, namely basis mismatch, non key-basis match, non-maximum signal intensity or detector efficiency, inconclusive outcome $b=00$, Bob will disclose all outcome results $b$.
    By doing so, they can estimate the statistics $f^{\{xyab\}}(\mu,\eta)$ and use Section \ref{sec:methods} to compute $q^{\{xya\}}(10|1), q^{\{xya\}}(01|1)$ for all relevant $x,y,a$ and then the achievable secure key rate the way we describe below.
    If the latter is positive, they proceed to step 5, otherwise they abort the protocol.

    \item \textbf{Post-processing}: For the remaining rounds, Alice and Bob apply suitable error correction and privacy amplification procedures to extract the secret key.
\end{enumerate}

The quantities $q^{\{xya\}}(kl |n)$ when $k+l=n=1$ are enough to characterise the proportion of events that can be analysed using conventional single photon security proof.
We show below how to relate them to the security of the qubit six-state protocol.

We define the qubit detection probability for a basis match as:
\begin{equation}
p^{x=y}_\text{det} = \sum\limits_{a\in \mathcal{A}}\sum\limits_{k+l=1} \Pr(a) q^{\{xya\}}(k,l |1) \Big(1 - r_k^{\{0\}}(\eta_0)r_l^{\{0\}}(\eta_1) \Big)
\end{equation}
In other words, a qubit detection happens when there is only one photon in any arm of the receiver and any conclusive detection pattern occurs.

Let us assume that the detectors are labelled in a way that in the absence of noise and $x=y$, there could be photons arriving only on detector number $a$, according to Alice's bit.
This allows us to define an ideal detection, hence any other result would be an erroneous detection.
We can define the success rate for the match basis:
\begin{equation}
p^{x=y}_\text{suc} = \frac{1}{p^{x=y}_\text{det}}\sum\limits_{a\in \mathcal{A}} \Pr(a) q^{\{xya\}}(a\oplus1, a|1)\Big(1 - r_{a\oplus1}^{\{0\}}(\eta_0)r_a^{\{0\}}(\eta_1) \Big)
\end{equation}
where $\oplus$ is the addition modulo $2$ and the error rate is defined as $p^{x=y}_\text{err} = 1 - p^{x=y}_\text{suc}$.

For simplicity and to relate to existing notation in the literature, we denote:
\begin{align}
p^{x=y=0}_\text{err} = e_Z\\
p^{x=y=1}_\text{err} = e_X\\
p^{x=y=2}_\text{err} = e_Y
\end{align}

We rephrase the results summarised in Appendix A of Ref.~\cite{scarani_security_2009} that is if we denote $\lambda = (\lambda_0, \lambda_1,\lambda_2, \lambda_3)$ to be the unique solution of:
\begin{equation}
\begin{matrix}
\lambda_0 &+& \lambda_1 &&&& &=& 1-e_Z\\
\lambda_0 &&&+&\lambda_2&& &=& 1-e_X\\
\lambda_0 &&&&&+&\lambda_3&=& 1-e_Y\\
\lambda_0 &+& \lambda_1 &+& \lambda_2 &+& \lambda_3 &=& 1\\
\end{matrix}
\end{equation}
and $H(\lambda) = -\sum\limits_{i=0}^3 \lambda_i \log_2(\lambda_i)$ the Shannon entropy, then the conditional entropy on Alice's bit value given Eve's side information is as follows:
\begin{equation}
H(A|E) \geq 1 - \Big( H(\lambda) - h_2(e_Z) \Big)
\end{equation}

We also define the signal detection rate and error rate for the key basis $0$:
\begin{align}
Q(\mu,\eta) &= 1 - \sum\limits_{a\in \mathcal{A}} \Pr(a) f^{\{x=y=0, a, b=00\}}(\mu,\eta)\\
E(\mu,\eta) &= 1 - \frac{1}{Q(\mu,\eta)} \sum\limits_{a\in \mathcal{A}} \Pr(a) f^{\{x=y=0, a, b=(a\oplus1, a)\}}(\mu,\eta)
\end{align}

Eventually, the achievable secure key rate against collective attacks and without sifting prefactor reads:
\begin{equation}
K \geq p_0(\mu)q(00|0)\big(1- r_0^{\{0\}}(\eta_0)r_0^{\{0\}}(\eta_1)\big) + p_1(\mu)p^{x=y=0}_\text{det} \Big( 1 - \big( H(\lambda) - h_2(e_Z) \big) \Big) - Q(\mu, \eta)h_2\big(E(\mu,\eta) \big)\\
\end{equation}

In the asymptotic analysis, we can always consider an efficient implementation where the maximum intensity $\mu$ and efficiency $\eta$ are used most of the time.
Otherwise, the framework allows to extract key from any ---key state--- intensity and efficiency setting.
This would be interesting for example in combination with a fast passive decoy states scheme \cite{mauerer_quantum_2007, curty_non-poissonian_2009, xu_improvement_2009, curty_passive_2010, zhang_simple_2018} since no extra sifting would be required.

We simulated an implementation of this protocol with one common threshold detector: an InGaAs SPAD with dark count rate $10^{-6}$ and efficiency $100\%$ (the actual finite efficiency is included in the total loss attributed to the channel) ; and a fixed channel error rate of $5\%$ in Fig.~\ref{fig:qkd_plot-1}. We compared the upper bound on the channel error rate given by our method and the one given in Ref.~\cite{ma_practical_2005} in Fig.~\ref{fig:qkd_plot-2}.

\end{document}